 \documentclass[preprint,12pt]{elsarticle}

\usepackage{amssymb}
\usepackage{comment}

\usepackage{lineno}

\journal{Nuclear Physics A}

\def\ffc{F_{\rm C}}
\def\ffm{F_{\rm M}}
\def\het{^3{\rm He}}
\def\htt{^3{\rm H}}

\def\hbq{{\hat Q}}
\def\hq{{\hat q}}
\def\hp{{\hat p}}
\def\hbp{{\hat P}}

\def\hko{{\hat k}_1}
\def\hkd{{\hat k}_2}
\def\hkt{{\hat k}_3}
\def\hkop{{\hat k}'_1}

\begin{document}

\begin{frontmatter}



\title{Trinucleon form factors with relativistic multirank separable kernels}

\author[label1]{Serge  Bondarenko}\ead{bondarenko@jinr.ru}
\author[label1]{Valery Burov}
\author[label1]{Sergey Yurev}
\address[label1]{BLTP, Joint Institute for Nuclear Research, Dubna, 141980, Russia }

\begin{abstract}
This paper studies elastic electron-trinucleon scattering 
in the relativistic impulse approximation. The amplitudes for 
a trinucleon have been obtained by solving the relativistic 
generalization of the Faddeev equations with a multirank  
separable kernel of the nucleon-nucleon interactions.
The static approximation and additional relativistic corrections for 
the trinucleon electromagnetic form factors have been calculated
for the momentum transfer squared up to 50~fm$^{-2}$.
\end{abstract}




\end{frontmatter}



\section{Introduction}
\label{intro}

In the previous article~\cite{Bondarenko:2020baw} we used the solutions of the 
Bethe-Salpeter-Faddeev equation (BSF) with the rank-one Yamaguchi and 
Tabakin separable kernels of the nucleon-nucleon (NN) interactions
to calculate the $\het$ charge form factor ($\ffc$).
Two approximations were considered: the static approximation (SA) and 
relativistic corrections (RC). 
The choice of the simple rank-one kernel did not allow 
us to reproduce the diffraction minimum of the form factor $\ffc$.

Authors of~\cite{Rupp:1991th} have found that the calculations with multirank 
separable NN kernel of interactions in the SA can produce the first diffraction 
minimum of the form factor $\ffc$ although at the higher value 
of the momentum transfer. 
But nevertheless, this article has focused us on the idea
that the RC can move the first diffraction minimum closer to the one experimentally observed. 

In the current paper we have considered several multirank kernels: purely phenomenological 
covariant generalization of the nonrelativistic (NR) Graz-II~\cite{Mathelitsch:1981mr} 
potential and the derived one from separable approximation of the 
Paris~\cite{Haidenbauer:1984dz,Haidenbauer:1985zz} NN potential. 
The parameters of the NR separable kernels 
were refitted in~\cite{Rupp:1991th}
to use them in the Bethe-Salpeter approach.

Today there are several relativistic approaches known to calculate trinucleon 
form factors (see, review~\cite{Marcucci:2015rca}).
The first one is the conventional approach: it views the nucleus as made up 
of nucleons interacting among themselves (via two- and many-body 
realistic potentials), and with external electromagnetic (EM) fields
(via one- and many-body currents), including relativistic corrections.
The second approach is the chiral effective field theory (ChEFT)
which uses the methods of quantum field theory
with Lagrangian including the nucleon, pion and photon fields
to obtain one- and many-body EM currents.
The third approach applies the covariant spectator theory (CST) to construct 
the three-nucleon wave function.
Within the framework of the latest approach, the corresponding covariant 
equation is solved by means of the one-boson-exchange kernel of the interactions
including several mesons.

One should also mention the light-front relativistic Hamiltonian dynamics (RHD)
which takes the three-body forces~\cite{Baroncini:2008eu} into account.

The main difference of the BSF formalism from the approaches mentioned above is 
that all the nucleons of the trinucleon BSF amplitude are off-mass-shell and 
there are two more free variables $p_0, q_0$ used in the calculations. 
These variables introduce additional difficulties in obtaining the solution 
of the BSF equations and calculating the form factors, namely, 
the peculiarities in the $p_0, q_0$ complex plane.

One of the important physical problems in the study of
the trinucleon form factors is to describe the right 
location of the first diffraction minimum.
For the $\het$ and $\htt$ charge and magnetic form factors
the result for the conventional and ChEFT approaches 
successfully reproduces the measured form factors up 
to momentum transfers squared $t \simeq 9$~fm$^{-2}$.
The ChEFT results seem to underestimate the experimental data beyond 
$t \gtrsim 9$~fm$^{-2}$, in particular, they predict the
form factor zeros at significantly lower values of $t$.
On the other hand, the CST calculations are limited to the approximation 
which omits the interaction current contribution. Therefore, 
only the isoscalar magnetic contribution should be compared to the
experimental data. For this isoscalar magnetic observable the agreement is good.

The article follows the ideas given in~\cite{Bondarenko:2020baw} 
where the rank-one kernels were applied.
In contrast to it, in this work the calculations with the multirank 
relativistic separable kernels of NN interactions Graz-II and Paris 
have been performed. The results of the calculations better agree with 
the experimental data than the results obtained
with the rank-one kernels -- Yamagychi and Tabakin.

The paper is organized as follows: Sec. 2 gives the expressions for  
the trinucleon EM form factors, 
Sec. 3 defines the static approximation and
relativistic corrections, 
Sec. 4 discusses the calculations and results,
and, finally, Sec. 5 gives the conclusion.

\section{Trinucleon form factors}
As a system with one-half spin, the EM current of $\het$ or $\htt$
can be parameterized by two elastic form factors: charge (electric) $\ffc$ and
magnetic $\ffm$ (see, for example,~\cite{Sitenko:1971}).
In the calculations below, we apply a straightforward relativistic generalization
of the NR expressions for the $\het$ and $\htt$ charge and magnetic 
form factors~\cite{Sitenko:1971,Schiff:1964zz,Gibson:1965zza,Rupp:1987cw}:
\begin{eqnarray}
&&2\ffc(\het) =  (2F^p_{\rm C} + F^n_{\rm C} )F_1  - \frac23(F^p_{\rm C} - F^n_{\rm C})F_2, 
\label{F_He_ch}
\\
&&F_{C}(\htt)  =  (2F^n_{\rm C} + F^p_{\rm C} )F_1  +  \frac23(F^p_{\rm C} - F^n_{\rm C})F_2, 
\label{F_H_ch}
\\
&&\mu(\het)F_{\rm M}(\het) =   \mu_nF^n_{\rm M} F_1  + \frac23(\mu_nF^n_{\rm M} + \mu_pF^p_{\rm M})F_2 + \frac43 (F^p_{\rm M} - F^n_{\rm M}) F_4,
	\label{F_He_mg}
\\
&&\mu(\htt)F_{\rm M}(\htt)  =   \mu_pF^p_{\rm M} F_1  + \frac23(\mu_nF^n_{\rm M} + \mu_pF^p_{\rm M})F_2 + \frac43 (F^n_{\rm M} - F^p_{\rm M}) F_4, 
\label{F_H_mg}
\end{eqnarray}
where $F^{p,n}_{\rm C}$, $F^{p,n}_{\rm M}$ are the charge and magnetic form factors 
of the proton and neutron, $\mu(\het)$, $\mu(\htt)$, $\mu_p$, $\mu_n$  
are  magnetic moments of the $\het$, $\htt$, proton and neutron,
respectively.

The functions $F_{1-4}$ can be expressed in terms of the wave
functions of the trinucleon which can be represented as linear combinations 
of the BSF equation solutions with different spin-isospin 
states~\cite{Bondarenko:2020baw,Sitenko:1971}:
\begin{eqnarray}
&&F_{1}(\hbq) =  \int d^4{\hp} \int d^4{\hq}\, G_1'(\hkop)\, G_1(\hko)\, G_2(\hkd)\, G_3(\hkt)\,
\sum_{i=1}^3\Psi^*_i(\hp,\hq;\hbp)\, \Psi_i(\hp,\hq';\hbp'),
\nonumber\\
&&F_{2}(\hbq) = -3 \int d^4 {\hp} \int d^4 {\hq}\, G_1'(\hkop)\, G_1(\hko)\, G_2(\hkd)\, G_3(\hkt)\,
\Psi^*_1(\hp,\hq;\hbp)\, \Psi_2(\hp,\hq';\hbp'),
\nonumber\\
\nonumber\\
&&F_{4}(\hbq) = \int d^4 {\hp} \int d^4 {\hq}\, G_1'(\hkop)\, G_1(\hko)\, G_2(\hkd)\, G_3(\hkt)\,
\Psi^*_3(\hp,\hq;\hbp)\, \Psi_3(\hp,\hq';\hbp'),
\label{f123}
\end{eqnarray}

In our previous work we performed the calculations for the rank-one interaction 
kernels~\cite{Bondarenko:2020baw,Bondarenko:2019gcd}.            
There we also analyzed the influence of the type of nucleon form factors on the $\het$
form factor~\cite{Bondarenko:2019gcd}. In the current paper we deal with the multirank 
separable kernels Graz-II (with $p_d$=4,5,6\%), Paris-1 and Paris-2.

To calculate the functions $F_{1-4}$, it is convenient to use the Breit 
reference system. The solutions of the BSF equation, however,
have been found in the c.m. (rest) frames of the corresponding trinucleon.
To relate the Breit and initial (final) particle c.m. frames, the Lorentz transformations 
should be applied to the four-momenta. 

Thus, the arguments of the initial and final particle wave functions and 
propagators were expressed in terms of the momenta calculated in the 
corresponding c.m. frames and related to each other using the Lorentz 
transformations. 

\section{Static approximation and relativistic corrections}

Below we remind the general approximations and corrections
(see detailed formulae in~\cite{Bondarenko:2020baw}).
Since the solutions of the BSF equations have been obtained 
in the Euclidean space~\cite{Bondarenko:2018xoq}
and are known only for real values of $q_4$ and $p_4$, the simplest way to
calculate $F_{1-4}$ is to apply the Wick rotation procedure 
$p_0 \to ip_4, q_0 \to iq_4$, if it is possible.
Thus, one needs to investigate the analytic structure of the 
integrand on the complex-valued variables $p_0,q_0$.
The location of the variable $p_0$ singularities allows one to apply 
the Wick rotation procedure, although this is not justified
for the variable $q_0$ in general.

It is convenient to start with the so-called static approximation.
It assumes that all the terms in the Lorentz transformations 
proportional to $\eta$ are canceled and the relativistic covariance 
for the EM current matrix element is violated. Analyzing integrands 
of $F_{1-4}$ in this case, one can see that the poles on $q_0$ do not cross
the imaginary $q_0$ axis and always stay in the second and fourth quadrants.
Therefore, the Wick rotation procedure $q_0 \to iq_4$ can be applied.

To recover the relativistic covariance, we 
consider the relativistic corrections to SA.
They consist of three parts.
First, we calculate the Lorentz boost in the one-particle 
propagator $G_1'(k_1')$ arguments, which gives a boost contribution (BC).
Second, we take into account a simple pole on $q_0$, which gives 
an additional term in integrals -- a pole contribution (PC).
Third, we compute the Lorentz boost of the arguments
of the final trinucleon wave function by carrying out the first 
term of the Taylor series expansion contribution (EC)
on the parameter $\eta$.

\section{Results and discussion}

As in the case of the rank-one kernel, we have used numerical solutions 
for the trinucleon amplitudes obtained by solving 
the system of homogeneous integral BSF  equations by means 
of the Gaussian quadratures.
The solutions mentioned above were interpolated to the $(q_4,q)$ points
of integration to perform multiple integration in equations $F_{1-4}$
using the Vegas algorithm of the Monte-Carlo integrator.
The stability of the result was tested by changing the
$(q_4,q)$-meshes while solving the BSF equations. The best
mesh was taken as $N_1 = 35, N_2 = 25$ where $N_1 (N_2)$ stands
for the number of $q_4 (q)$ points. The multiple integrals were calculated with
the relative accuracy equal to 0.01.

\begin{figure}[ht]
	\centering
	\includegraphics[width=0.85\linewidth,angle=0]{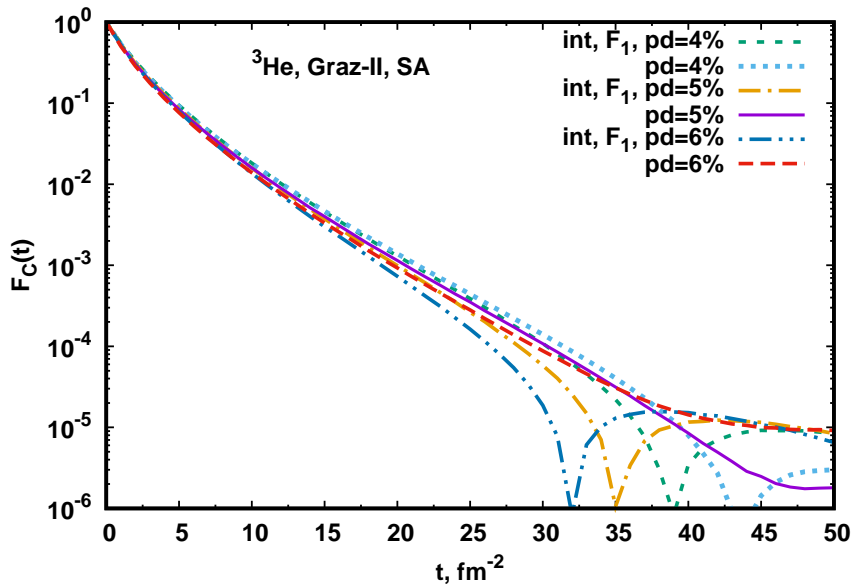}
	\caption{Static approximation for the charge form factor of $\het$ 
	as a function of $t$ for the multirank separable kernel Graz-II 
	at different values of the probability of the two-nucleon $D$-state $p_d$ = 4,5,6\%.
    Complete calculations (the dotted line -- for $p_d$=4\%, the solid line -- for $p_d$=5\% and the long-dashed line -- for $p_d$=6\%) 
    are compared to the~\cite{Rupp:1991th} calculations within certain approximations 
    (the dashed line -- for $p_d$=4\%, the dashed-dotted line -- for $p_d$=5\% and the dashed-dotted-dotted line -- for $p_d$=6\%).	   
}
	\label{fig1}
\end{figure}

\begin{figure}[ht]
	\centering
	\includegraphics[width=0.85\linewidth,angle=0]{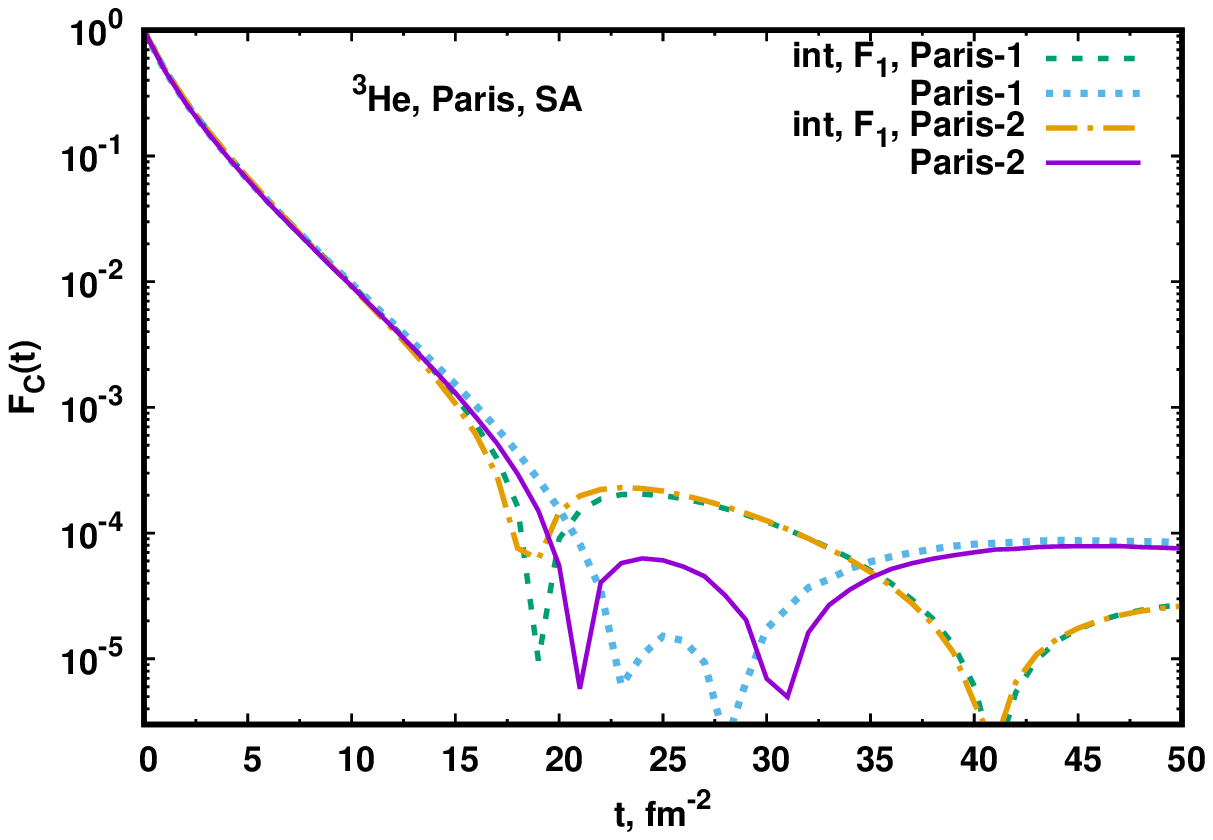}
	\caption{Static approximation for the charge form factor of $\het$ 
	as a function of $t$ for the multirank separable kernel Paris.
    Complete calculations (the dotted line -- for Paris-1 and the solid line -- for Paris-2) 
    are compared to the~\cite{Rupp:1991th} calculations within certain approximation 
    (the dashed line -- for Paris-1, the dashed-dotted line -- for Paris-2).  
}
	\label{fig1a}
\end{figure}

It should be emphasized that the considered separable covariant Graz-II kernels
were obtained from the purely phenomenological
nonrelativistic Graz-II potential, while the covariant Paris-1(2)
kernels were received from the NR realistic one-boson-exchange Paris
model.

In this section we use the dipole fit for nucleon form factors
if not stated otherwise.

\begin{figure}[ht]
	\centering
	\includegraphics[width=0.85\linewidth,angle=0]{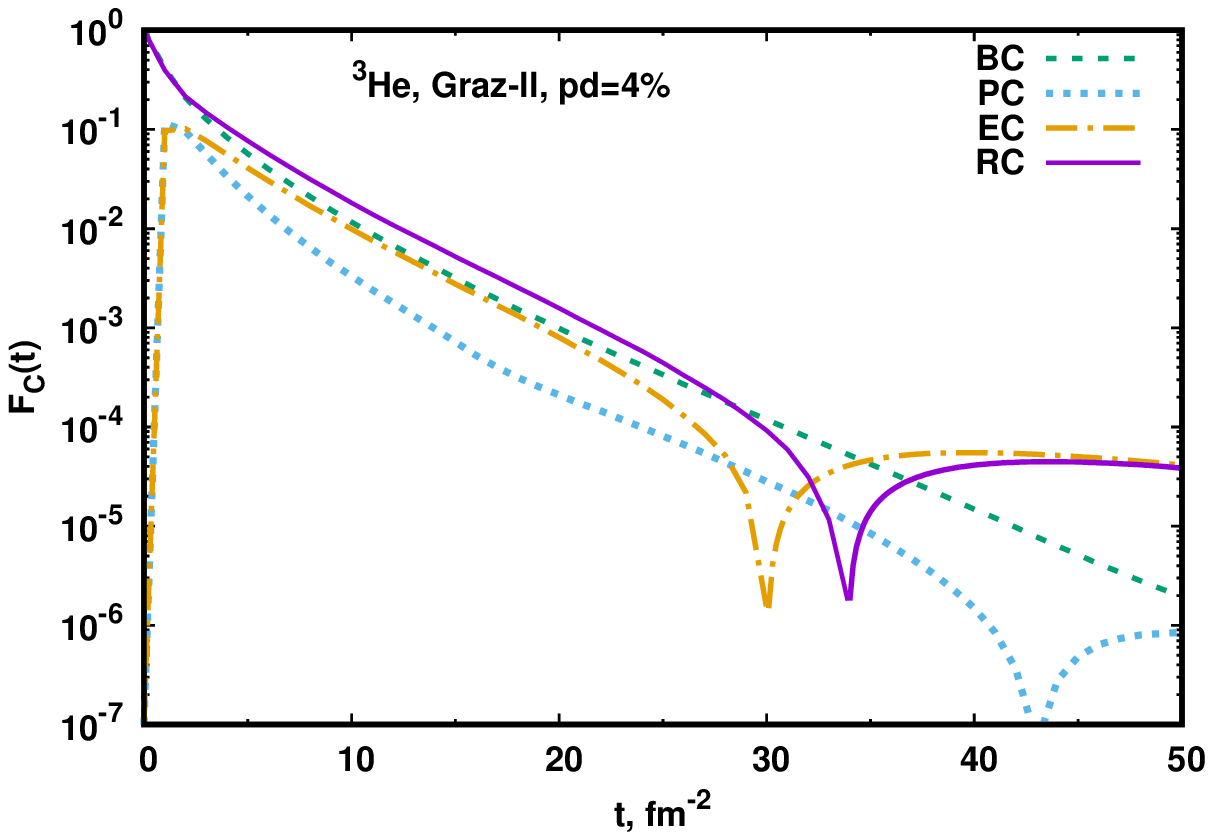}	
	\caption{Parts of contribution to relativistic corrections for the $\het$ charge form factor
		as a function of $t$
		for the multirank separable kernel Graz-II 
		at the probability of the two-nucleon $D$-state $p_d = 4\%$.
		The solid line displays the sum of all considered 
		relativistic corrections,
		the dashed line -- Lorentz transformations of arguments of a propagator, 
		the dotted line displays the contribution of the simple pole of the one-particle propagator,
		and the dashed-dotted line -- Lorentz transformations of the arguments of the final particle
		wave function.
		}
	\label{fig2}
\end{figure}

\begin{figure}[ht]
	\centering
	\includegraphics[width=0.85\linewidth,angle=0]{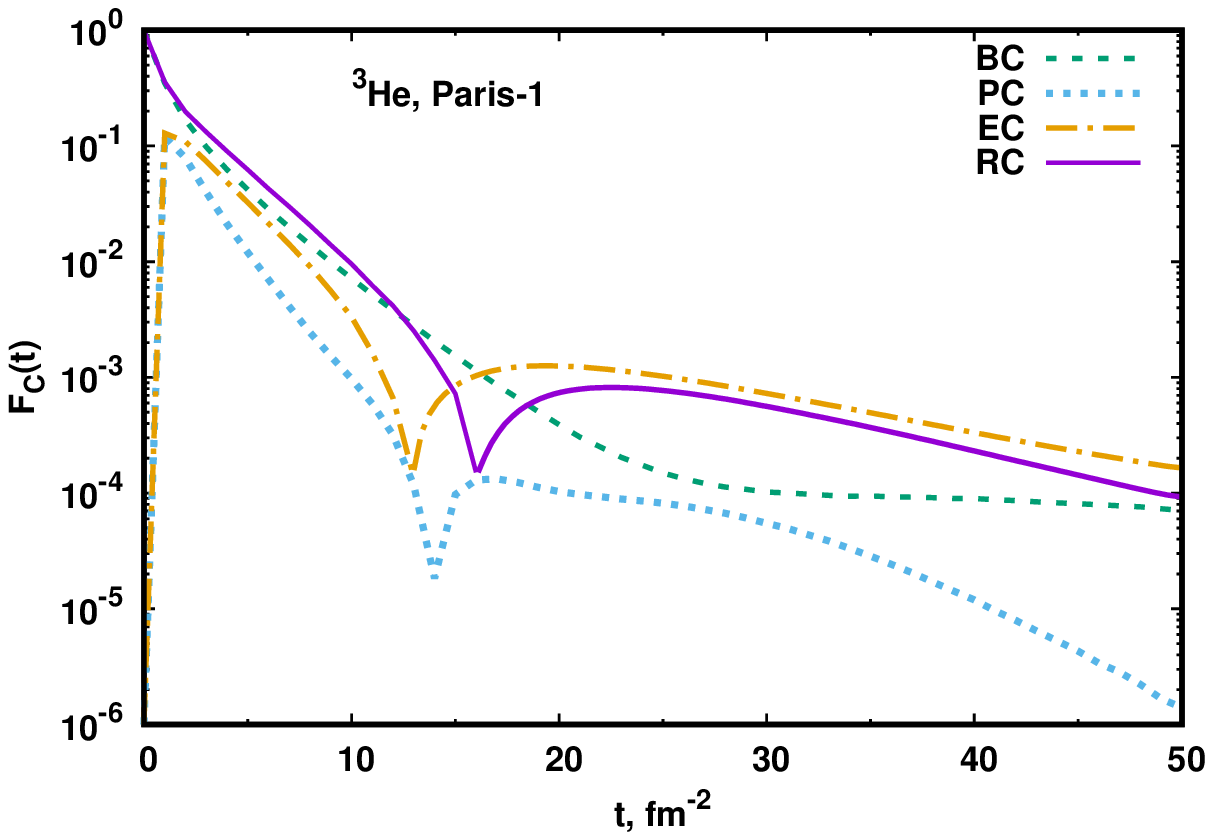}	
	\caption{The same as in Fig.~\ref{fig2} for
		the multirank separable kernel Paris-1.
		}
	\label{fig2a}
\end{figure}

Now we discuss the comparison of the SA results with the ones obtained in paper~\cite{Rupp:1991th}
within the following approximation: the functions $\psi_{2,3}$ which depend on angular integration 
$x=\cos\theta_{qp}$ were averaged numerically while the integration in the propagators was performed analytically.
The angular averaging leads to $\psi_{2} = \psi_{3} \sim \psi_1$ which 
consequently gives the following condition: $v_2 \sim 0$. To imitate the~\cite{Rupp:1991th} calculations, we have used the analytical expressions for the propagators and taken into account the $F_1$ function only with the integral calculated numerically without averaging on the $x$ variable. The obtained results are very similar to those shown in~\cite{Rupp:1991th} 
but they differ from the complete calculations. The reason is that in complete integration 
the function $v_2$ is not equal to zero and gives a noticeable contribution into function $F_2$ which dominates in the region of the form factor zero.

Figure~\ref{fig1} shows the result of the calculations of the $\het$ charge form factor 
for the cases of complete integrations and the approximation from~\cite{Rupp:1991th} 
for the kernel Graz-II.
As mentioned above, there is a big difference between these two assumptions.
Figure~\ref{fig1a} illustrates the same as in Figure~\ref{fig1} but for the Paris-1 and Paris-2 kernels.

Now we discuss the contribution of the RC to the $\het$ and $\htt$ charge and magnetic form factors.
As it is shown in~\cite{Bondarenko:2020baw} neither the SA nor RC 
calculated with the rank-one NN kernel give the diffraction minimum in the form factors.
The static approximation within the multirank kernels produces the minimum/zero in the form factors
which is shifted to the higher momentum transfer squared region in comparison to the experimental data.
We expect the RC to shift the minimum/zero closer to the position of the experimental data.

Figure~\ref{fig2} shows the partial contributions of the RC
to the $\het$ charge form factor, namely, the BC, PC and EC
for the Graz-II multirank kernel with $p_d=4\%$.
Figure~\ref{fig2a} illustrates the same for the Paris-1 kernel.
At the low values of $t$=1-2~fm$^{-2}$ the PC and EC are small
due to the small corresponding value of variable $\eta$
and the RC are defined by the BC.
Starting approximately from $t = 5$~fm$^{-2}$ the values of the BC and EC 
are equal to each other
and their sum makes the major contribution to RC.
As it is seen from the Figures above the form factor zero at value
$t = 33$~fm$^{-2}$ for the Graz-II kernel 
(at $t = 16$~fm$^{-2}$ for Paris-1)
is defined by zero of the EC at $t = 30$~fm$^{-2}$ 
for Graz-II (at $t = 12$~fm$^{-2}$ for Paris-1)
and shifted by the positive contribution of the BC.
In the region after the form factor zero
the BC sharply decreases and the RC are totally
determined by the EC. 
The PC is rather small in comparison to the BC and EC and does not play 
a significant role except for the case when the sum of the BC and EC
is close to the value of PC at small $t$ 
and near the form factor zero.

\begin{figure}[ht]
	\centering
	\includegraphics[width=0.85\linewidth,angle=0]{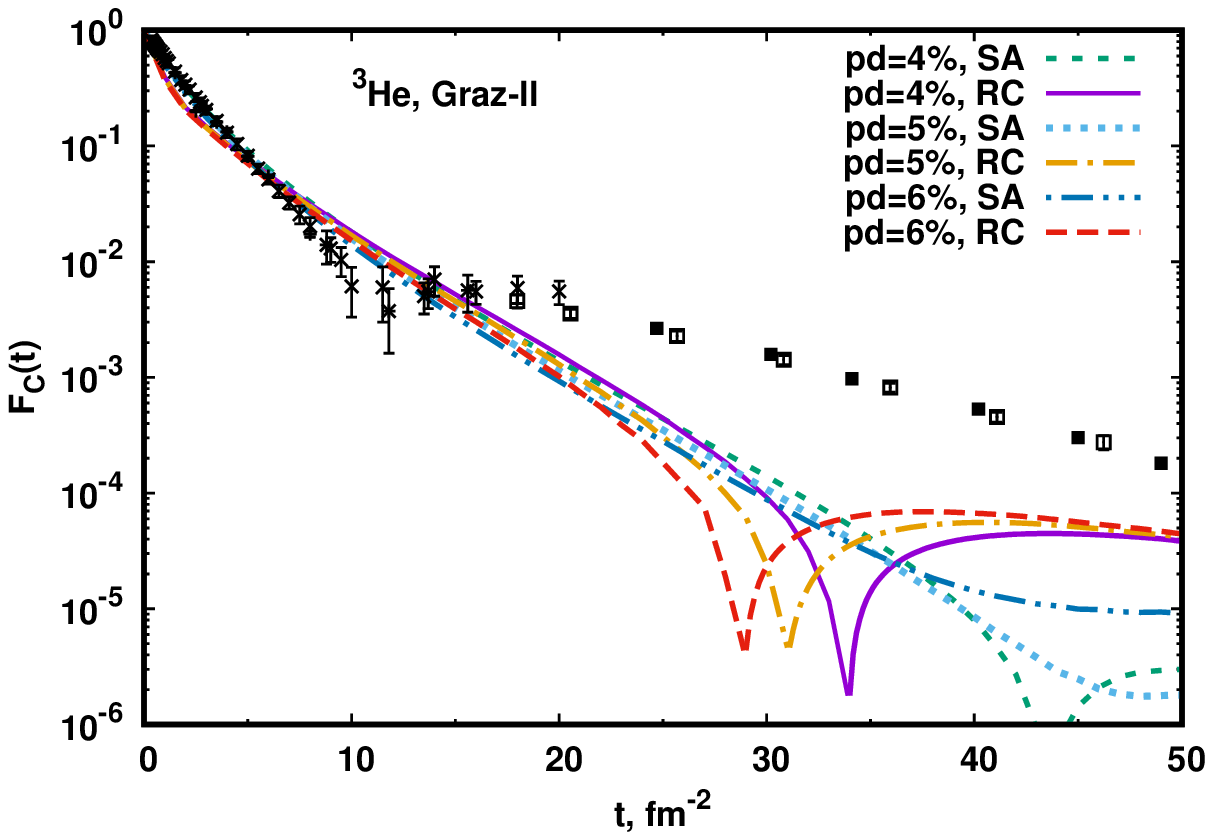}	
	\caption{Static approximation and relativistic corrections for the $\het$ charge form factor
		as a function of $t$
		for the multirank separable kernel Graz-II 
		at different values of the probability of the two-nucleon $D$-state $p_d = 4,5,6\%$.
		The dashed line -- SA for $p_d$=4\%,
		the dotted line -- SA for $p_d$=5\%,
		the dashed-dotted-dotted line -- SA for $p_d$=6\%,
        the solid line -- RC for $p_d$=4\%, 
        the dashed-dotted line -- RC for $p_d$=5\% 
        and the long-dashed line -- RC for $p_d$=6\%.
		The experimental data are taken from~\cite{Stanford_1965, Bernheim:1972az, Mccarthy:1977vd,
		Arnold:1978qs, Camsonne:2016ged}.
		}
	\label{fig3}
\end{figure}

Figure~\ref{fig3} shows the SA and RC for the $\het$ charge form factor
as a function of $t$ for the multirank separable kernel Graz-II 
at different values of the probability of the two-nucleon $D$-state 
$p_d = 4,5,6\%$.
For all the curves there is a diffraction minimum/zero except for the cases 
of the SA with $p_d = 5,6\%$ 
where the minimum without changing the sign is found. 
The relativistic corrections for all the values of $p_d$ shift zeros 
to the region of the smaller $t$ closer to the experimental data.
The difference between the RC and experimental zeros is about 18-23~fm$^{-2}$. 

\begin{figure}[ht]
    \centering
	\includegraphics[width=0.85\linewidth,angle=0]{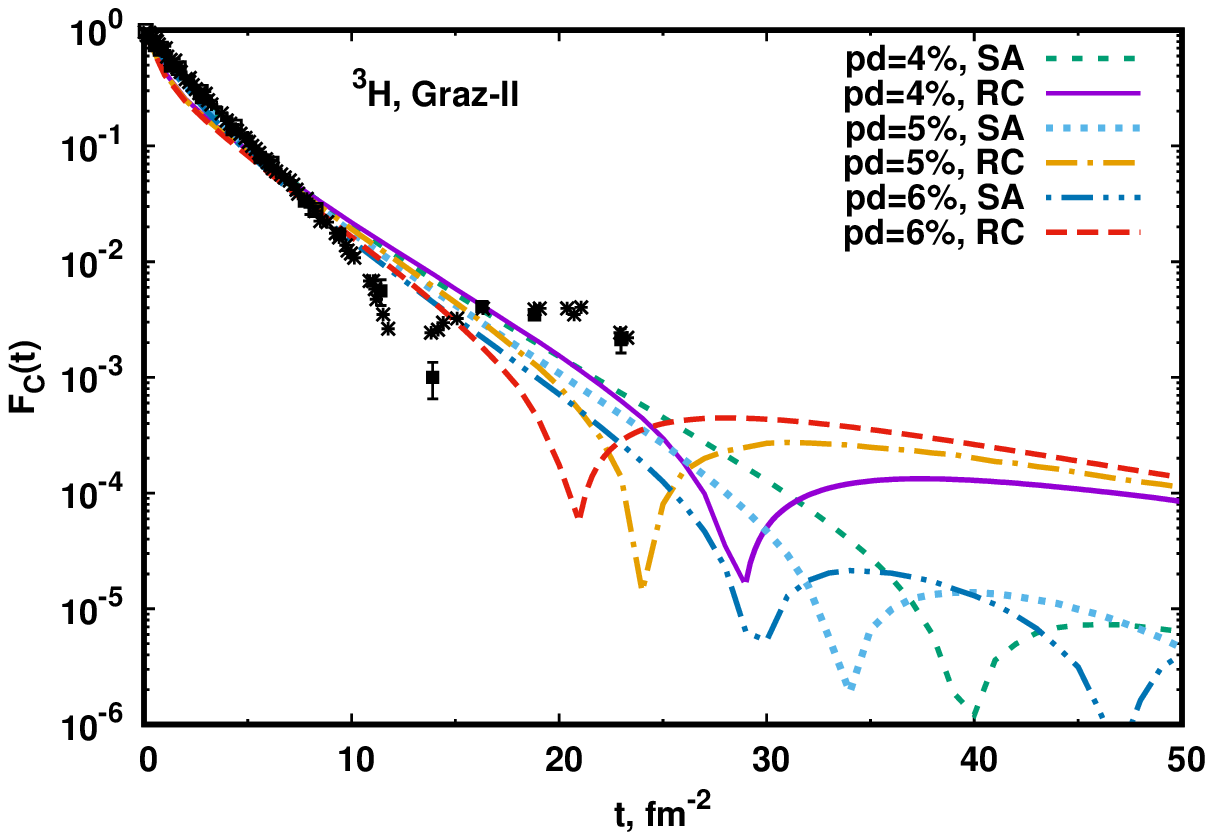}
	\caption{The same as in Fig.~{\ref{fig3}} but for the $\htt$ charge form factor.
		The experimental data are taken from~\cite{Stanford_1965, MIT-BATES_1983-3h, SACLAY_1985,
		BATES_1987, Slac_1994}.
	}
	\label{fig4}
\end{figure}

Figure~\ref{fig4} presents the SA and RC for the $\htt$ charge form factor.
In this case the SA gives the first form factor zero at about
40~fm$^{-2}$ for $p_d$=4\%, 
34~fm$^{-2}$ for $p_d$=5\% and 
29~fm$^{-2}$ for $p_d$=6\%.
For all $p_d$ values the RC shift the results closer 
to the experimental data of the diffraction minima
in comparison to the SA.
The difference between the RC and experimental zeros is less than in the $\het$ charge 
form factor and is about 9-15~fm$^{-2}$.

\begin{figure}[ht]
	\centering
	\includegraphics[width=0.85\linewidth,angle=0]{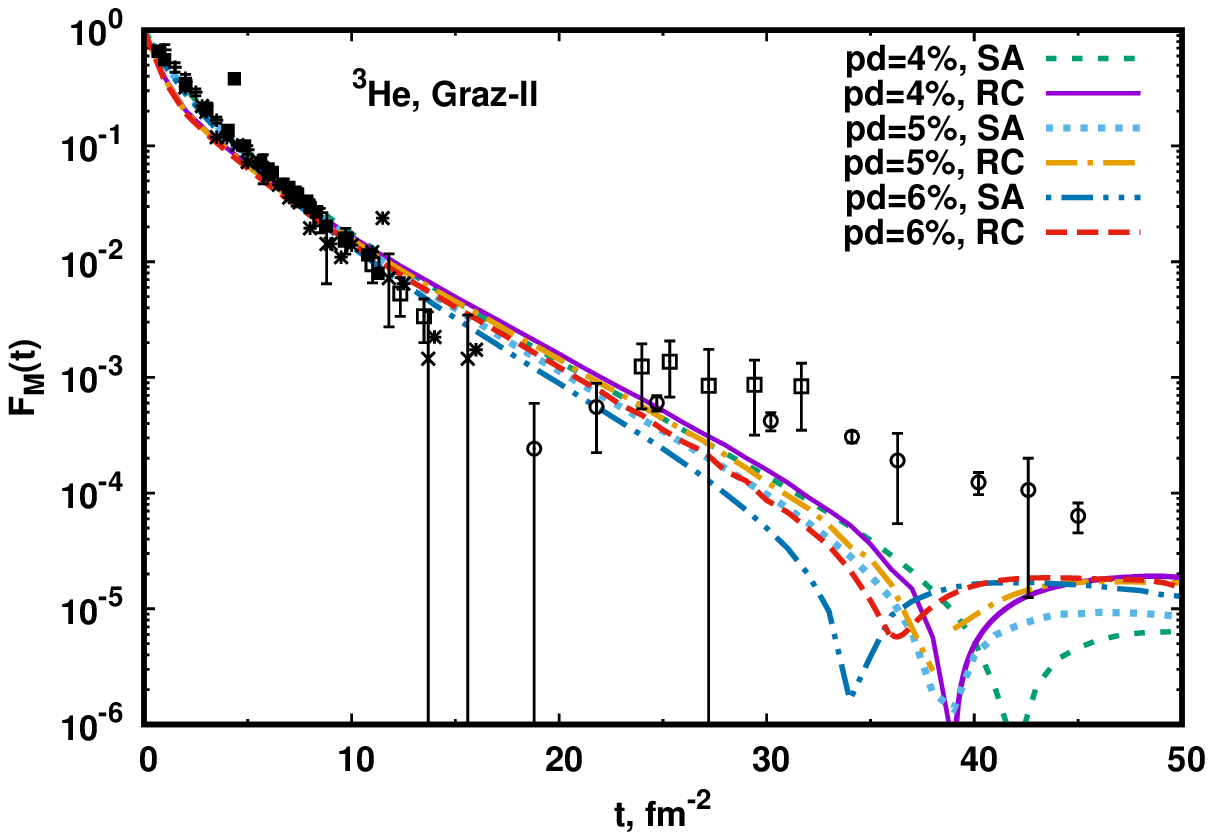}
	\caption{The same as in Fig.~{\ref{fig3}} but for the $\het$ magnetic form factor.
		The experimental data are taken from~\cite{Stanford_1965, Bernheim:1972az, Mccarthy:1977vd, Camsonne:2016ged,
		SACLAY_1982, MIT-BATES_1983, MIT-BATES_2001}.
		}
	\label{fig5}
\end{figure}

Figure~\ref{fig5} shows the SA and RC for the $\het$ magnetic form factor.
The results are similar to the $\het$ charge form factor.
However, the difference between the RC and experimental zeros is larger than in the $\het$ charge form factor and is about 15-17~fm$^{-2}$.

\begin{figure}[ht]
	\centering
	\includegraphics[width=0.85\linewidth,angle=0]{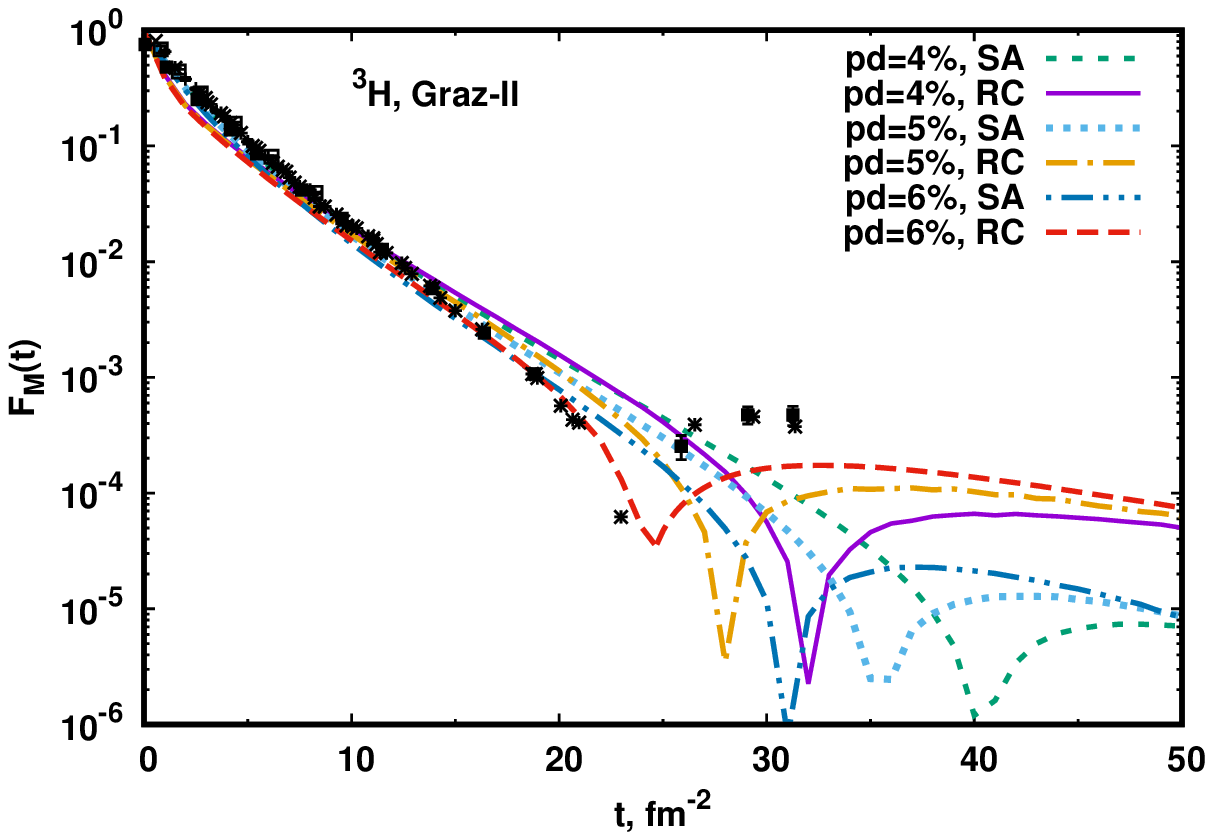}	
	\caption{The same as in Fig.~{\ref{fig3}} but for the $\htt$ magnetic form factor.
		The experimental data are taken from~\cite{Stanford_1965, MIT-BATES_1983-3h, SACLAY_1985, 
		BATES_1987, Slac_1994}.
	}
	\label{fig6}
\end{figure}

Figure~\ref{fig6} displays the SA and RC for the $\htt$ magnetic form factor. 
The results are similar to the $\htt$ charge form factor (Fig.~\ref{fig3}). 
In this case the difference between 
the RC and experimental zeros is less than in the $\htt$ charge form factor
and is about 2-10~fm$^{-2}$ for $p_d$=4,5\%.
The relativistic corrections for $p_d = 6\%$ 
agree with the experimental data zero and their behaviour is rather well
described.

\begin{figure}[ht]
	\centering
	\includegraphics[width=0.85\linewidth,angle=0]{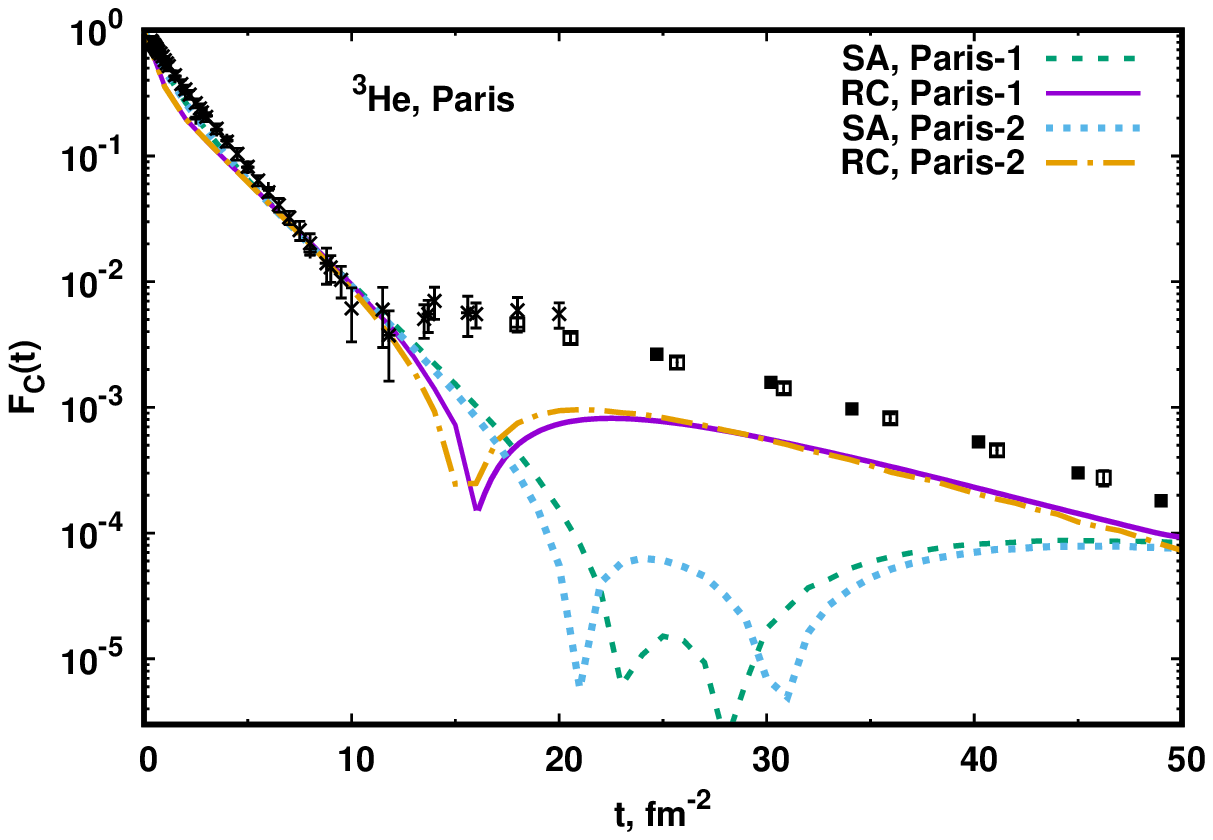}	
	\caption{Static approximation and relativistic corrections for the $\het$ charge form factor
		as a function of $t$
		for the multirank separable kernel Paris.
				The Dashed line -- SA for Paris-1,
				the dotted line -- SA for Paris-2,
				the solid line -- RC for Paris-1,
			and the dashed-dotted line -- RC for Paris-2.
	The experimental data coincide with Fig.~\ref{fig3}.
		}
	\label{fig3a}
\end{figure}

\begin{figure}[ht]
\centering
	\includegraphics[width=0.85\linewidth,angle=0]{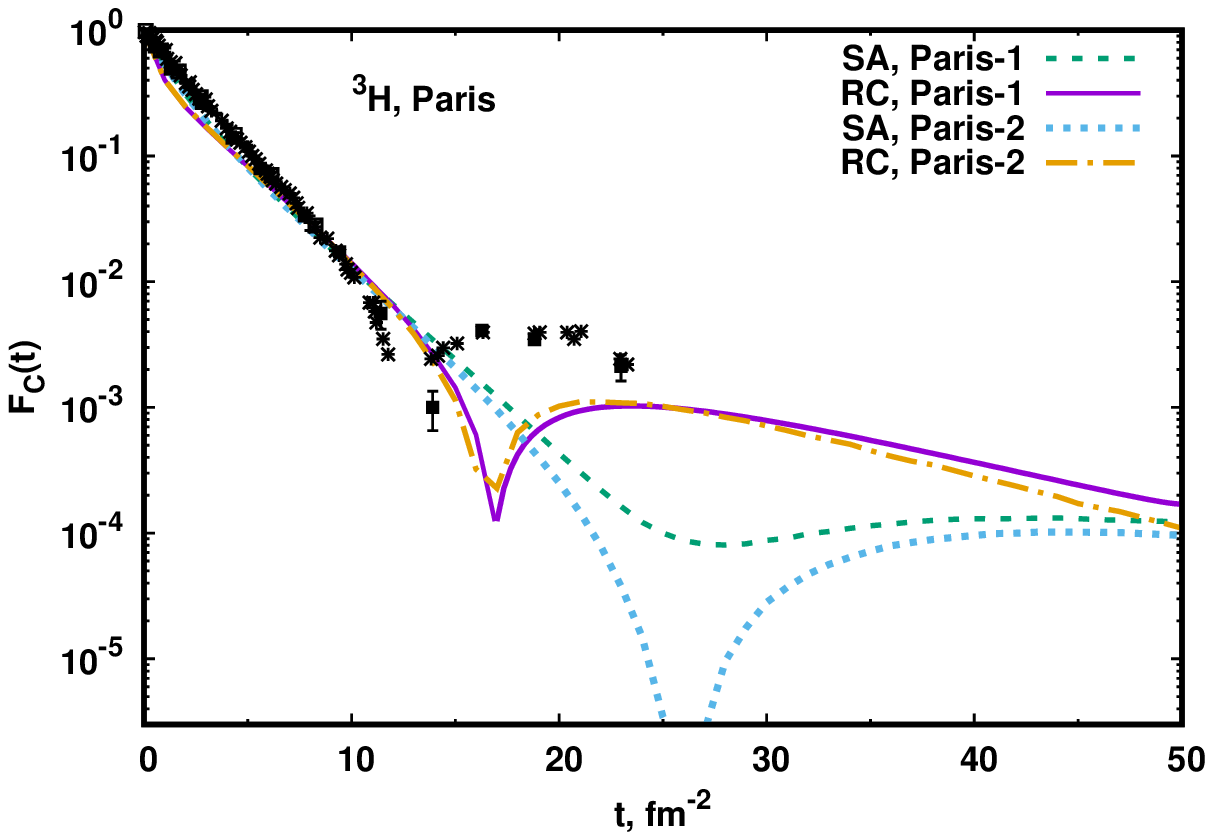}
	\caption{The same as in Fig.~{\ref{fig3a}} but for the $\htt$ charge form factor.
	The experimental data coincide with Fig.~\ref{fig4}.
	}
	\label{fig4a}
\end{figure}

\begin{figure}[ht]
	\centering
	\includegraphics[width=0.85\linewidth,angle=0]{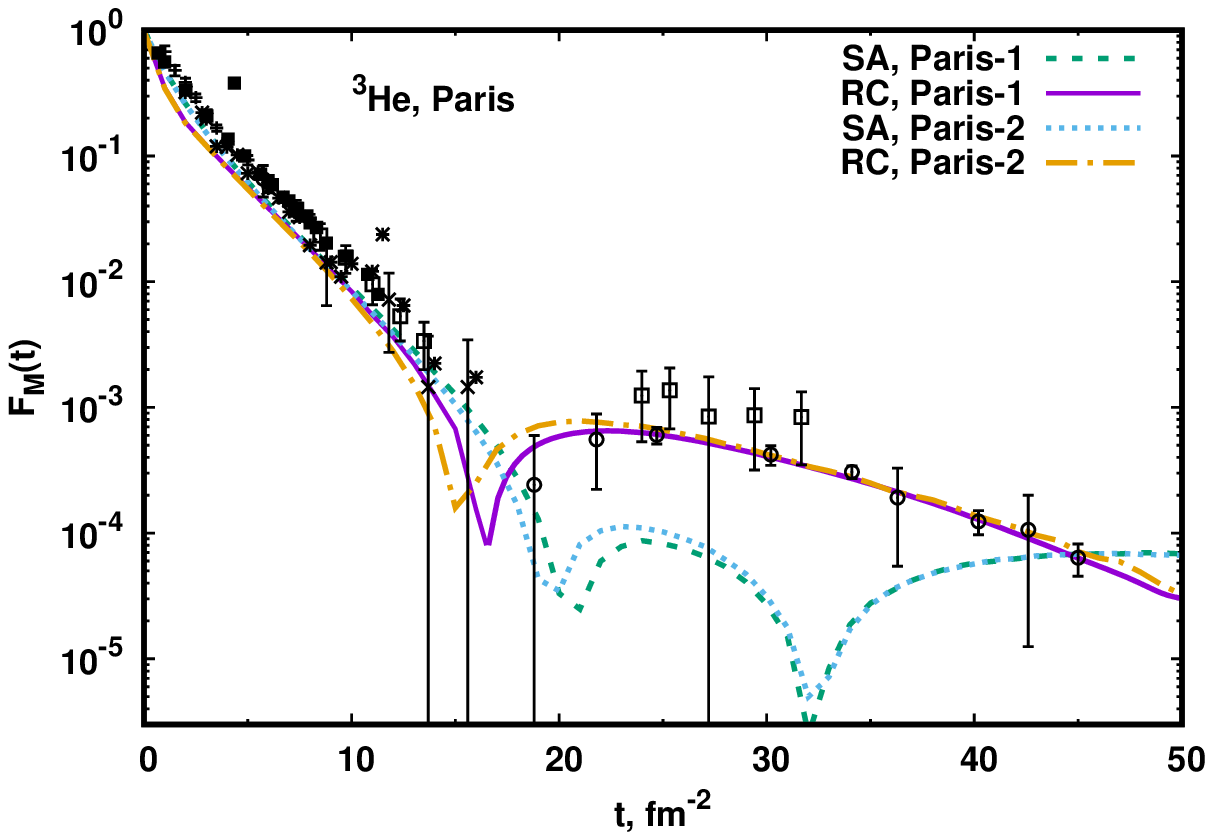}
	\caption{The same as in Fig.~{\ref{fig3a}} but for the $\het$ magnetic form factor.
	The experimental data coincide with Fig.~\ref{fig5}.
		}
	\label{fig5a}
\end{figure}

\begin{figure}[ht]
	\centering
	\includegraphics[width=0.85\linewidth,angle=0]{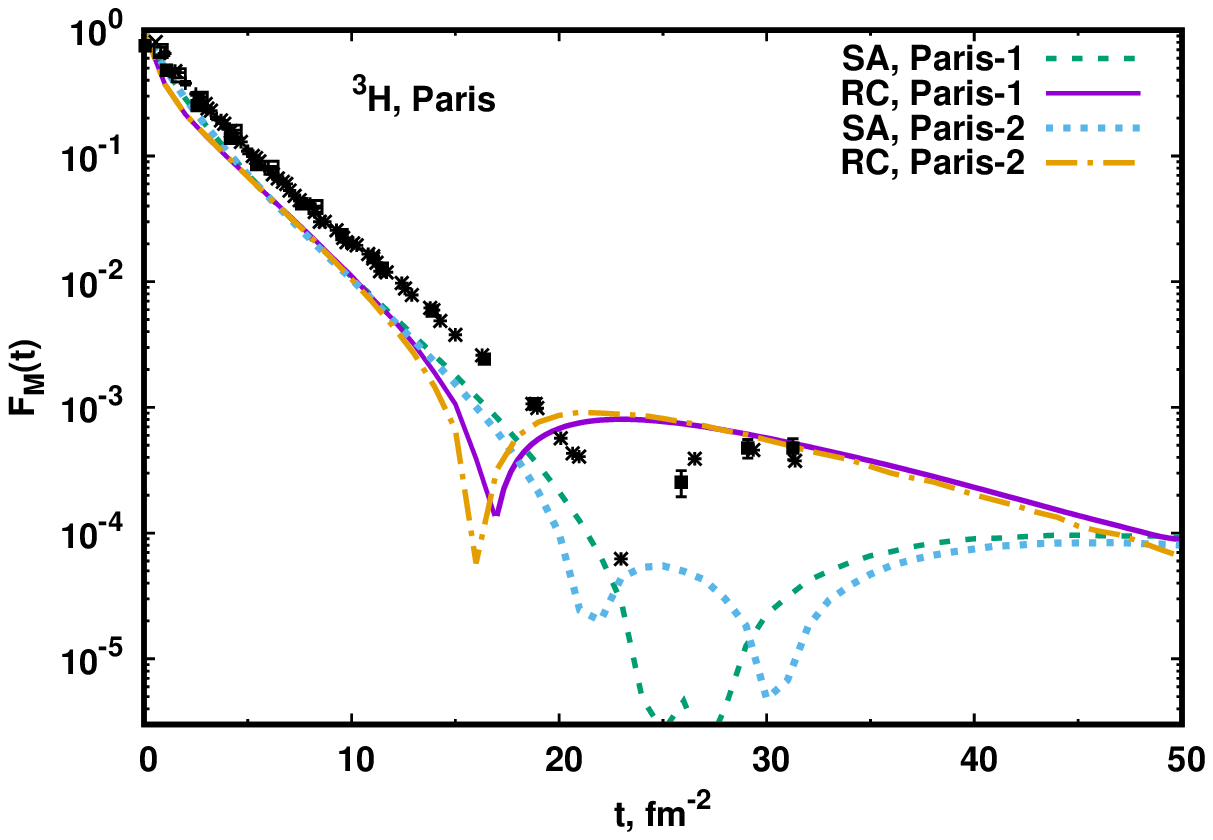}	
	\caption{The same as in Fig.~{\ref{fig3a}} but for the $\htt$ magnetic form factor.
	The experimental data coincide with Fig.~\ref{fig6}.
	}
	\label{fig6a}
\end{figure}

The results for the $\het$ and $\htt$ charge and magnetic form factors with the
Paris-1 and Paris-2 kernels are presented in Figures~\ref{fig3a}--\ref{fig6a}.
The calculation results indicate that in this case as well as in the case of the Graz-II kernel 
the RC shift the curves towards the experimental data. 
The difference between the data and calculation results 
is generally smaller than for the Graz-II kernel. 
Figure~\ref{fig3a} demonstrates that for the $\het$ charge form factor 
the RC are rather close to the data zero and
the difference is small -- about 5(6)~fm$^{-2}$ for Paris-1(2) kernels. 
Opposite to the $\het$ charge form factor
the zero position and behaviour of the magnetic
one in the region $t$=16-50 ~fm$^{-2}$ 
are in rather good agreement with the experimental data
(Figures~\ref{fig5a}) while the $\htt$ magnetic form factor the
RC have zeros shifted to the smaller values of $t$
in comparison to the data.

It should be stressed that in most cases of the above-mentioned multirank
kernels (except the Paris-1(2) for $\het$ and $\htt$ magnetic form factors)
the RC also increase the value of the form factors by an order of magnitude 0.5-1.5 at the value of $t=50$~fm$^{-2}$ in comparison to the SA calculations.

\begin{figure}[ht]
	\centering
	\includegraphics[width=0.85\linewidth,angle=0]{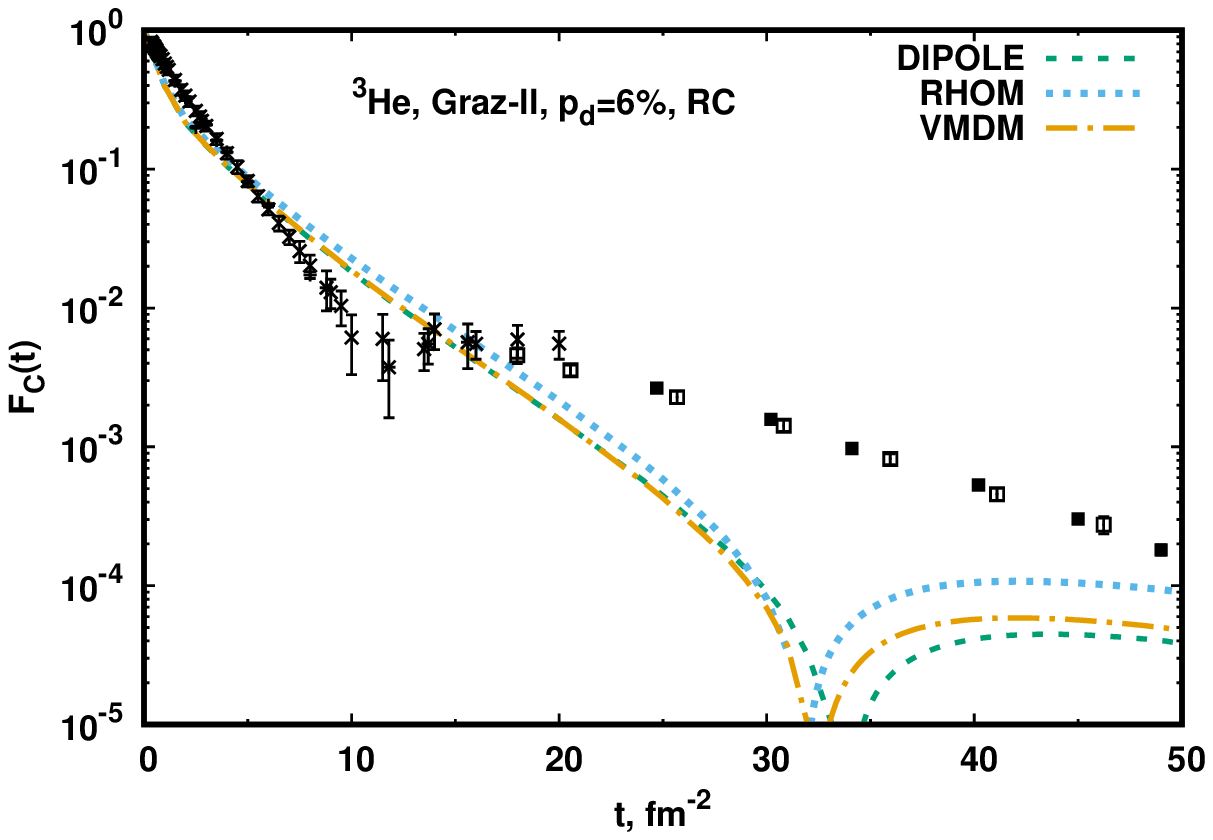}	
	\caption{Relativistic corrections to the $\het$ charge form factor for Graz-II 
	($p_d=4\%$) kernel with three models of the EM nucleon form factors:
	the dashed line - DIPOLE,
	the dotted line - RHOM,
	the dashed-dotted line - VMDM.
	The experimental data coincide with Fig.~\ref{fig6}.
	}
	\label{fig7}
\end{figure}

\begin{figure}[ht]
	\centering
	\includegraphics[width=0.85\linewidth,angle=0]{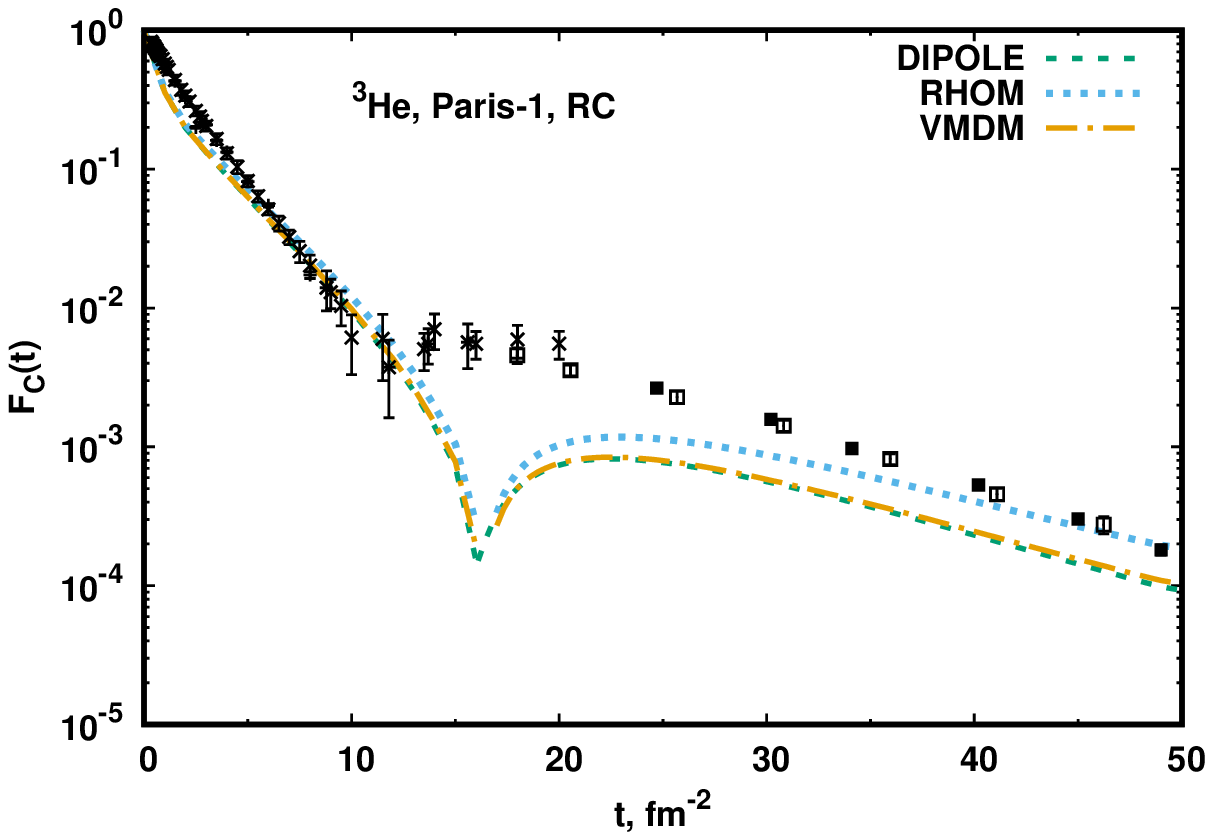}	
	\caption{The same as in Fig.~\ref{fig7}
	but for the Paris-1 kernel.
	}
	\label{fig8}
\end{figure}

Finally, we consider the contribution of different models of the
nucleon EM form factors: the dipole model (DIPOLE), relativistic harmonic
oscillator model (RHOM)~\cite{Burov:1993ia} and vector meson dominance model
(VMDM)~\cite{Bijker:2004yu}. The VMDM has used the latest experimental 
data to refit the parameters.
The results for the above three models are given
for the $\het$
charge form factor
in Fig.~\ref{fig7} for Graz-II and
in Fig.~\ref{fig8} for Paris kernels.
The Figures demonstrate that the models change the magnitudes of the form factors 
by a factor of 1.5-2 at $t=50$~fm$^{-2}$ and also shift slightly the form factor zero.
The main effect of the above changes happens due to the non-zero electric neutron form 
factor $G^n_E$ in the RHOM and VMDM.

We have to emphasize that the considered RC are nothing but the restoration of the covariance 
of the trinuleon EM current matrix element
although at the first order of the Taylor series expansion
of the final trinucleon wave function arguments.
The effect of RC is large and important as it is seen from the results.
These RC change the static approximation results in the proper direction. 
We have calculated
the relativistic impulse approximation (RIA) which
takes into account only three contributions from 
the individual photon-nucleon diagrams
and also considers the nucleon EM form factors on-mass-shell (as CIA discussed above).
The RIA obviously violates the gauge invariance of the reaction. 
To recover this invariance, one should introduce the corresponding EM interaction currents 
which together with RIA provide the gauge invariance.
Hopefully, this contribution will improve the 
agreement between the RC and experimental data.

\section{Summary}

In this paper the BSF equations solved for trinucleon have been 
used to calculate the trinucleon electromagnetic form factors. 
The expressions for the form factors have been obtained by a straightforward
relativistic generalization of the NR expressions.
The multiple integration has been performed by means of the Monte Carlo algorithm.
The Lorentz transformations 
of the propagator arguments and the final trinucleon wave function have been taken into account while calculating.

The obtained solutions with multirank kernels Graz-II and Paris-1(2) unlike 
the rank-one kernels have given the diffraction minimum. 
It is important to stress that the strong dependence on the type of the NN interaction kernel has been found in the calculations.

Two approximations have been considered: the static approximation and 
relativistic corrections.
The relativistic corrections provide the diffraction minimum in the form 
factors and move it in the proper direction
even if it does not appear in the static approximation. 

For the Graz-II with $p_d=6\%$ and Paris-1(2) kernels
a good agreement has found for the position of the diffraction minimum 
in $\ffm$ of $\htt$ and of $\het$, respectively.

Finally, we can also conclude that taking into account the EM interaction
currents will be justified to provide the gauge 
invariance and further progress in the study of
elastic electron-trinucleon scattering.

\bibliographystyle{elsarticle-num}

\end{document}